\documentclass[dvips]{article}
\usepackage{icrctc07}

\title{A Spectacular VHE Gamma-Ray Outburst from PKS 2155-304 in 2006}
\shorttitle{A Spectacular VHE $\gamma$-ray Outburst from PKS\,2155$-$304}
\authors{W.\,Benbow$^{1}$, 
C.\,Boisson$^{2}$, 
L.\,Costamante$^{1}$, 
O.\,de\,Jager$^{3}$,
G.\,Dubus$^{4}$,
D.\,Emmanoulopoulos$^{5}$, 
B.\,Giebels$^{6}$,
S.\,Pita$^{7}$,
M.\,Punch$^{7}$,
C.\,Raubenheimer$^{3}$,
M.\,Raue$^{8}$,
H.\,Sol$^{2}$,
and S.\,Wagner$^{5}$
for the HESS Collaboration
}
\shortauthors{W.\,Benbow et al.}
\afiliations{$^1$ Max-Planck-Institut f\"ur Kernphysik, Heidelberg, Germany; 
$^2$ LUTH, UMR 8102 du CNRS, Observatoire de Paris, Section de Meudon, France; 
$^3$ Unit for Space Physics, North-West University, Potchefstroom, South Africa;
$^4$ Laboratoire d'Astrophysique de Grenoble, INSU/CNRS, Universit\'e Joseph Fourier, France;
$^5$ Landessternwarte, Heidelberg, Germany;
$^6$ LLR, CNRS/IN2P3, Ecole Polytechnique, Palaiseau, France;
$^7$ Astrophysique et Cosmologie (APC), Paris, France;
$^8$ Universit\"at Hamburg, Institut fuer Experimentalphysik, Germany
}
\email{Wystan.Benbow@mpi-hd.mpg.de}

\abstract{

Since 2002 the VHE ($>$100 GeV) $\gamma$-ray flux of
the high-frequency peaked BL\,Lac PKS\,2155$-$304
has been monitored with the High Energy Stereoscopic System (HESS).  An extreme
$\gamma$-ray outburst was detected in the early hours of July 28, 2006
(MJD 53944).  The average flux above 200 GeV observed during this outburst is
$\sim$7 times the flux observed from the Crab Nebula above the same threshold.  
Peak fluxes are measured with one-minute time scale resolution at more than twice
this average value. Variability is seen up to $\sim$600 s in the
Fourier power spectrum, and well-resolved bursts varying on time
scales of $\sim$200 seconds are observed. There are no strong
indications for spectral variability within the data.  Assuming the
emission region has a size comparable to the Schwarzschild radius of a $\sim$$
10^9\,M_\odot$ black hole, Doppler factors greater than 100 are
required to accommodate the observed variability time scales.
}

\begin{document}
\maketitle

\section{Introduction}

In the Southern Hemisphere, PKS\,2155$-$304 (redshift $z=0.116$) is
generally the brightest blazar at VHE energies, and is probably the
best-studied at all wavelengths.  The VHE flux observed
\cite{HESS_2155A} from PKS\,2155$-$304 is typically of the order
$\sim$15\% of the Crab Nebula flux above 200 GeV.  The highest flux
previously measured in one night is approximately four times this
value and clear VHE-flux variability has been observed on daily time
scales. The most rapid flux variability measured for this source is
25\,min~\cite{HESS_2155B}, occurring at X-ray energies.  The fastest
variation published from any blazar, at any wavelength, is an event lasting
$\sim$800\,s where the X-ray flux from Mkn\,501 varied by 30\%
\cite{MKN501_dispute}\footnote{Xue \& Cui~\cite{MKN501_dispute}
also demonstrate that a 60\% X-ray flux increase in $\sim$200\,s observed
\cite{MKN501_flare} from Mkn\,501 is likely an artifact.}, 
while at VHE energies doubling time scales as fast as
$\sim$15 minutes have been observed from Mkn 421 \cite{Gaidos_Mkn421}.

As part of the normal HESS observation program
the flux from known VHE AGN is monitored regularly
to search for bright flares.  During the July 2006 dark period,
the average VHE flux observed by HESS from PKS\,2155$-$304 was
more than ten times its typical value.  In particular,
an extremely bright flare of PKS\,2155$-$304 was 
observed in the early hours of July 28, 2006 (MJD 53944).
This contribution focuses solely on this particular flare,
which is described in more detail in \cite{2155_letter}.

\section{Results from MJD 53944}
\label{sect:results}

A total of three observation runs ($\sim$28 min each) were
taken on PKS\,2155$-$304 in the early hours of MJD 53944.  These data
entirely pass the standard HESS data-quality selection criteria,
yielding an exposure of 1.32\,h live time at a mean zenith angle of
13$^{\circ}$.  The analysis method is described in \cite{2155_letter}.
The observed excess is 11771 events (168$\sigma$), corresponding to a
rate of $\sim$2.5\,Hz.  This is the first time the
detected VHE  $\gamma$-ray rate has exceeded 1\,Hz.

\subsection{Flux Variability}

\begin{figure}[t]
  \centering
  \includegraphics[width=0.45\textwidth]{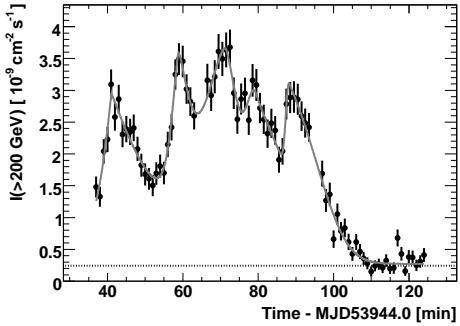}
  \caption{I($>$200 GeV) observed from PKS\,2155$-$304 binned
    in 1-minute intervals.  The horizontal line represents
    I($>$200 GeV) observed \cite{hess_crab}
    from the Crab Nebula.  The curve is the fit
    to these data of the superposition of five bursts (see text) and a
    constant flux.\label{flux_lc_1min}}
\end{figure}

\begin{figure}
  \centering
  \includegraphics[width=0.45\textwidth]{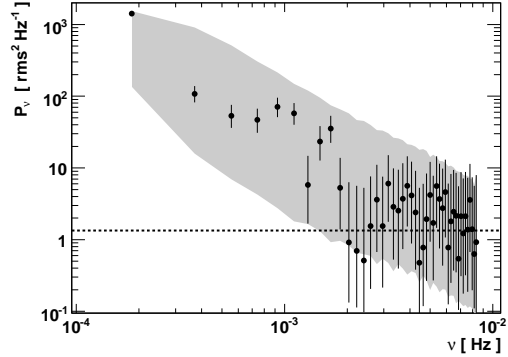}
  \caption{The Fourier power spectrum of the
    light curve and associated measurement error. The grey shaded area
    corresponds to the 90\% confidence interval for a light curve with a power-law
    Fourier spectrum $P_{\nu}\propto \nu^{-2}$. The horizontal line is the
    average noise level. \label{fourier_power}}
\end{figure}

The average integral flux above 200 GeV observed from PKS\,2155$-$304
is I($>$200 GeV) = 
(1.72$\pm$$0.05_{\rm stat}$$\pm$$0.34_{\rm syst}$)$\,\times\,$10$^{-9}$\,cm$^{-2}$\,s$^{-1}$, 
equivalent to $\sim$7 times the I($>$200 GeV) observed 
from the Crab Nebula (I$_{\mathrm{Crab}}$, \cite{hess_crab}).  Figure~\ref{flux_lc_1min}
shows I($>$200 GeV), binned in one-minute intervals, versus time.  The
fluxes in this light curve range from 0.65 I$_{\mathrm{Crab}}$ to 15.1
I$_{\mathrm{Crab}}$, and their fractional root mean square (rms)
variability amplitude \cite{rms_noise_ref} 
is F$_{\rm var}=0.58\pm0.03$.  This is $\sim$2
times higher than archival X-ray 
variability \cite{zhang1999,zhang2005}. 
The Fourier power spectrum
calculated from Figure~\ref{flux_lc_1min}
is shown in Figure~\ref{fourier_power}. 
There is power
significantly above the measurement noise level up to $1.6 \times
10^{-3}\,{\rm Hz}$ ($600\,{\rm s}$).  
The power spectrum derived from the data is
compatible with a light curve generated by a stochastic process with a
power-law Fourier spectrum of index -2. An index of -1 produces too
much power at high frequencies and is rejected. These power spectra are
remarkably similar to those derived in
X-rays \cite{zhang1999} from the same source.

Figure~\ref{flux_lc_1min} clearly contains substructures
with even shorter rise and decay time scales than found 
in the Fourier analysis.  Therefore, the light curve is considered as
consisting of a series of bursts, which is common for AGN and
$\gamma$-ray bursts (GRBs).  To characterize these bursts, 
the ``generalized Gaussian'' shape from
Norris et al.~\cite{norris} is used,
where the burst intensity is described by: ${\rm I}(t) = A \exp [
-(|t-t_{\rm max}|/\sigma_{\rm r,d})^\kappa]$, where $t_{\rm max}$ is
the time of the burst's maximum intensity (A); $\sigma_{\rm r}$ and
$\sigma_{\rm d}$ are the rise ($t<t_{\rm max}$) and decay ($t>t_{\rm
max}$) time constants, respectively; and $\kappa$ is a measure of the
burst's sharpness.  The rise and decay times, from half to maximum
amplitude, are $\tau_{r,d}=[\ln 2]^{1/\kappa}\sigma_{r,d}$. 
Five significant bursts were found with a peak finding tool based on 
a Markov chain algorithm \cite{mor02}. 
The data are well fit\footnote{All parameters are left free in the fit.}
by a function consisting of a superposition of an identical
number of bursts plus a constant signal.   
The best fit has a $\chi^2$ probability of
20\% and the fit parameters are shown in
Table~\ref{burst_info}. Interestingly, there is a marginal trend for
$\kappa$ to increase with subsequent bursts, 
making them less sharp, as the flare
progresses, which could imply the bursts are not stochastic.
The $\kappa$ values 
are close to the bulk of those found by Norris et al.~\cite{norris}, 
but the time scales measured here are two orders of
magnitude larger.

\begin{table}
\caption{The results of the best $\chi^2$ fit of the superposition
of five bursts and a constant to the
data shown in Figure~\ref{flux_lc_1min}\label{burst_info}.}
\centering
\begin{tabular}{cccc}
\\
\hline\hline
          \noalign{\smallskip}
$t_{\rm max}$  & $\tau_{\rm r}$& $\tau_{\rm d}$  & $\kappa$ \\
$[$min$]$ & [s] & [s] & \\
          \noalign{\smallskip}
          \hline
          \noalign{\smallskip}
	  41.0 & 173$\pm$28 & 610$\pm$129 & 1.07$\pm$0.20\\
	  58.8 & 116$\pm$53 & 178$\pm$146 & 1.43$\pm$0.83\\
	  71.3 & 404$\pm$219 & 269$\pm$158 & 1.59$\pm$0.42\\
	  79.5 & 178$\pm$55  & 657$\pm$268 & 2.01$\pm$0.87\\
	  88.3 & 67$\pm$44   & 620$\pm$75  & 2.44$\pm$0.41\\
          \noalign{\smallskip}
          \hline
\end{tabular}
\end{table}

During both the first two bursts there is clear doubling of the flux
within $\tau_{r}$.  Such doubling is sometimes used as a
characteristic time scale of flux variability.  For compatibility with
such estimators, the definition of doubling time, 
$T_2 = |{\rm I}_{ij} \Delta T / \Delta {\rm I}|$, from 
\cite{zhang1999} is also used\footnote{Only values of $T_2$ with 
less than 30\% uncertainty are considered.}.  Here, $\Delta T = T_j - T_i$, 
$\Delta {\rm I} = {\rm I}_j - {\rm I}_i$, $
{\rm {\rm I}}_{ij} = ({\rm I}_j + {\rm I}_i)/2$, with $T$ 
and I being the time and flux, respectively, of any pair
of points in the light curve. 
The fastest $T_2=224\pm60\,{\rm s}$ is
compatible with the fastest significant time scale found by the
Fourier transform. Averaging the five lowest $T_2$
values yields $330\pm40\,{\rm s}$.

\subsection{Spectral Analysis}

\begin{figure}
   \centering
      \includegraphics[width=0.45\textwidth]{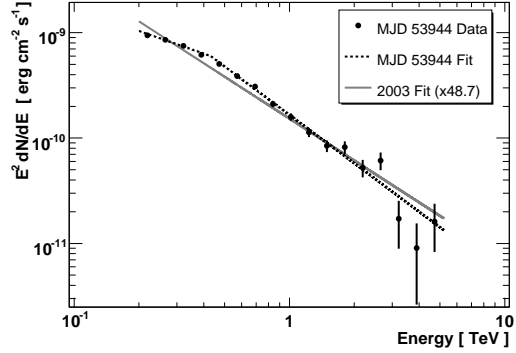}
\caption{The time-averaged spectrum observed
	from PKS\,2155$-$304 on MJD 53944.
	The dashed line is 
	the best $\chi^2$ fit of a broken power law to
	the data.  The solid line represents the fit to
	the time-averaged spectrum of PKS\,2155$-$304 
	from 2003 \cite{HESS_2155A} 
	scaled by 48.7.
\label{avg_spectrum}}
\end{figure}

Figure~\ref{avg_spectrum} shows the time-averaged photon 
spectrum for these data.  The data are well fit, $\chi^2=17.1$ for 13 
degrees of freedom (d.o.f.), by a broken power-law function with
$E_{\mathrm{B}}$\,=\,430$\pm$22$\pm$80 GeV,
$\Gamma_{1}$\,=\,2.71$\pm$0.06$\pm$0.10, 
and $\Gamma_{2}$\,=\,3.53$\pm$0.05$\pm$0.10. 
For each parameter, the two uncertainties are the 
statistical and systematic values, respectively.
The time-averaged spectrum ($\Gamma$\,=\,3.32) 
of PKS\,2155$-$304\ measured in 2003 \cite{HESS_2155A},
multiplied by the ratio (48.7) of I($>$200~GeV) from the 
respective data sets, is also shown in Figure~\ref{avg_spectrum}.
Despite a factor of $\sim$50 change in flux there 
is qualitatively little difference between the two spectra 
which is surprising. The lack of any strong ($\Delta \Gamma$\,$>$\,0.2)
temporal variability in the VHE spectrum within these data
(tested on time scales of 5, 10 and 28 minutes) is also surprising.

\section{Discussion}

It is very likely that the electromagnetic emission in blazars 
is generated in jets that are beamed 
and Doppler-boosted toward the observer. 
Superluminal expansions observed with VLBI \cite{piner}
provide evidence for moderate Doppler boosting in PKS\,2155$-$304.
Causality implies that $\gamma$-ray variability on a 
time scale $t_{\rm var}$, with a Doppler factor\footnote{With 
$\delta$ defined in the standard
way as $[\Gamma(1-\beta\cos\theta)]^{-1}$, where $\Gamma$ is the bulk
Lorentz factor of the plasma in the jet, $\beta = v/c$, and $\theta$
is the angle to the line of sight.} ($\delta$), is related to the
radius ($R$) of the emission zone by $R \leq ct_{\rm
var}\delta/(1+z)$. Conservatively using the best-determined rise time 
(i.e. $\tau_r$ with the smallest error) 
from Table~\ref{burst_info} for $t_{\rm var} = 173\pm28\,{\rm s}$ 
limits the size of the emission region
to $R\delta^{-1} \leq 4.65 \times 10^{12}$ cm $\leq 0.31$ AU.

The jets of blazars are believed to be powered by accretion onto a 
supermassive black hole (SMBH).  Thus accretion/ejection 
properties are usually presumed to scale with the Schwarzschild radius 
$R_{\rm S}$ of the SMBH, where $R_{\rm S} = 2GM/c^2$, which
is the smallest, most-natural size of the system
(see, e.g., \cite{Blandford}).
Expressing the size $R$ of the $\gamma$-ray emitting region 
in terms of $R_{\rm S}$, the variability time
scale limits its mass by $M \leq (c^3 t_{\rm var}\delta/2G(1+z)) R_{\rm S}/R 
\sim 1.6\times10^7 M_\odot \delta R_{\rm S} / R$. 
The reported host galaxy luminosity $M_R=-24.4$
(Table 3 in \cite{kotilainen}) would imply a SMBH mass of order
1$-$2$\times 10^9M_\odot$ \cite{bettoni2003}, and therefore,
$\delta\geq 60-120\,R/R_{\rm S}$.  Emission regions of only a 
few $R_{\rm S}$ would
require values of $\delta$ much greater than those typically
derived for blazars ($\delta$$\sim$10) and come close
to those used for GRBs, which would be a challenge to understand.  

Although the choice of a $\sim$3 minute variability time scale in this article is 
conservative, it is still the fastest ever
seen in a blazar, at any wavelength, and is almost an order of
magnitude smaller than previously observed from this object.
The variability is a factor of five times faster than 
VHE variability previously measured from Mkn 421 \cite{Gaidos_Mkn421}
and comparable to that reported from Mkn\,501 \cite{MAGIC_501}.
However, in terms of the 
light-crossing time of the Schwarzschild
radius, $R_{\rm S}/c$, the variability of PKS\,2155$-$304 is
more constraining by 
another factor\footnote{These factors assume
black hole masses of 
$10^{8.22} M_\odot$ and $10^{8.62} M_\odot$ for Mkn\,421 
and Mkn\,501, respectively \cite{BH_Mass}.}
of $\approx 6-12$ for Mkn\,421, and a factor of $\approx 2.5-5$ 
for  Mkn\,501. 

The light curve presented here is strongly oversampled,
allowing for the first time in the VHE regime a detailed 
statistical analysis of a flare, which shows remarkable similarity to
other longer duration events at X-ray energies. More detailed discussion
of this outburst can be found in \cite{2155_letter}, and the
event continues to be investigated with other statistical
techniques. As the sensitivity of VHE instruments continues to improve, 
it is likely that similar extreme flaring episodes will be more commonly
detected in the future. Similar flares will
only strengthen the conclusion that
either very large Doppler factors can
be present in AGN jets, or that the observed variability is not
connected to the central black hole.

\section{Acknowledgements}

The support of the Namibian authorities and of the University of Namibia
in facilitating the construction and operation of HESS is gratefully
acknowledged, as is the support by the German Ministry for Education and
Research (BMBF), the Max Planck Society, the French Ministry for Research,
the CNRS-IN2P3 and the Astroparticle Interdisciplinary Programme of the
CNRS, the U.K. Science and Technology Facilities Council (STFC),
the IPNP of the Charles University, the Polish Ministry of Science and
Higher Education, the South African Department of
Science and Technology and National Research Foundation, and by the
University of Namibia. We appreciate the excellent work of the technical
support staff in Berlin, Durham, Hamburg, Heidelberg, Palaiseau, Paris,
Saclay, and in Namibia in the construction and operation of the
equipment.


\end{document}